# Surface phase transitions in BiFeO$_3$ below room temperature

R. Jarrier[1,2], X. Marti[3], J. Herrero-Albillos[4,5], P. Ferrer[6,7], R. Haumont[1,2], P. Gemeiner[2], G. Geneste[2], P. Berthet[1], T. Schülli[8], P. Cvec[9], R. Blinc[9], Stanislaus S. Wong[10,11], Tae-Jin Park[10,12], M. Alexe[13], M. A. Carpenter[14], J.F. Scott[15], G. Catalan[16], B. Dkhil[2,*]

[1] *Laboratoire de Physico-Chimie de l'Etat Solide, ICMMO, CNRS-UMR 8182, Bâtiment 410 - Université Paris-Sud XI, 15 rue Georges Clémenceau 91405 Orsay Cedex, France*
[2] *Laboratoire Structures, Propriétés et Modélisation des Solides, CNRS-UMR8580, Ecole Centrale Paris, Grande Voie des Vignes, 92295 Chatenay-Malabry Cedex, France*
[3] *Department of Physics, Charles University, Prague*
[4] *Helmholtz-Zentrum Berlin für Materialien und Energie GMBH, Albert-Einstein-Straße 15, 12489 Berlin, Germany.*
[5] *Centro Universitario de la Defensa, Academia General Militar-Universidad de Zaragoza, 50090 Zaragoza, Spain*
[6] *SpLine (BM25), ESRF, Grenoble, France*
[7] *Instituto de Ciencia de Materiales de Madrid ICMM-CSIC, Madrid, Spain*
[8] *ESRF Beamline ID01, Grenoble, France.*
[9] *Jozef Stefan Institute, Jamova 39, Ljubljana 1000, Slovenia*
[10] *Department of Chemistry, State University of New York at Stony Brook, Stony Brook, NY 11794-3400, USA*
[11] *Also at Condensed Matter Physics and Materials Sciences Department, Brookhaven National Laboratory, Building 480, Upton, NY 11973, USA*
[12] *Also at Korea Atomic Energy Research Institute (KAERI), 989-111 Daedoek-daero, Yuseong, Daejeon, Korea 305-353*
[13] *Max Planck Institute for Microstructural Physics, Halle, Saale, Germany*
[14] *Dept. of Earth Sciences, University of Cambridge, Downing Street, Cambridge CB2 3EQ, U. K.*
[15] *Dept. Physics, Cavendish Lab., Cambridge Univ., Cambridge CB3 0HE, U. K.*
[16] *ICREA and CIN2 (CSIC-ICN), Universitat Autonoma de Barcelona, Bellaterra 08193, Spain*

Keywords: multiferroics, BiFeO$_3$, surface phase transition, low temperature anomalies

**PACS**: 76.50.+g 77.80.Bh 81.30.Dz

We combine a wide variety of experimental techniques to analyze two heretofore mysterious phase transitions in multiferroic bismuth ferrite at low temperature. Raman spectroscopy, resonant ultrasound spectroscopy, EPR, X-ray lattice constant measurements, conductivity and dielectric response, specific heat and pyroelectric data have been collected for two different types of samples: single crystals and, in order to maximize surface/volume ratio to enhance surface phase transition effects, BiFeO$_3$ nanotubes were also studied. The transition at T=140.3K is shown to be a surface phase transition, with an associated sharp change in lattice parameter and charge density at the surface. Meanwhile, the 201K anomaly appears to signal the onset of glassy behaviour.

## 1. Introduction

Bismuth ferrite BiFeO$_3$ (BFO) is one of the most popular research materials in condensed matter physics at present.[1, 2]. Despite the intense activity, however, there remain a number of unanswered questions concerning its structure and phase diagrams. From the beginning[3] a large number of phases were reported as a function of temperature, and more recently[4,5,6,7,8,9] more as a function of hydrostatic pressure. The high temperature end of the phase diagram is now resolved, based upon the neutron studies of Arnold et al.[10, 11] and involves an ambient rhombohedral R3c phase, a first-order transition to orthorhombic Pbnm near 1103K,[10] and an iso-symmetric Pbnm-Pbnm Mott-like metal-insulator transition near 1220K.[11] The powder neutron results show that the latter structure cannot be resolved by X-ray studies, because only the oxygen ions are significantly displaced; and it further shows that there are no high-T monoclinic or tetragonal phases, contrary to claims elsewhere.[12, 13, 14] In some non-powder thin-film specimens, a cubic Pm-3m phase is inferred a few degrees below the melting temperature of ca. 1225K[13] but it still remains to be confirmed.[16]

Although the high-temperature phases have been identified, questions remain about lower-temperature anomalies. The true nature and stability of its long-period (ca. 63 nm), incommensurate cycloidic spin structure has been controversial[17-25] and there are a number of cryogenic phase transitions whose origin has not been clarified. For example, anomalies near 140K and 201K[26, 27] have been interpreted as spin reorientation transitions,[27] analogous to those in orthoferrites such as ErFeO$_3$, and evidence has also been reported for spin-glass behaviour,[28, 29] with an Almeida-Thouless line (AT-line) terminating at 140K,[30] a clear separation of field-cooled and zero-field-cooled susceptibilities beginning at 230K, additional magnon light-scattering cross-section divergences near 90K and 230K, and a bump in the dielectric constant near 50K.[31-32] At ca. 30K, there are two anomalies: an extrapolated freezing temperature from a Vogel-Fulcher analysis of data (29.4K). These studies therefore indicate up to six cryogenic anomalies at temperatures at 30, 50, 90, 140, 201, and 230K. On the other hand, neutron diffraction experiments and other bulk-sensitive probes, such as single crystal magnetometry, show no indication of any magnetic transition, with the spin cycloid seemingly unaltered all the way down to 4K.[18, 25] There is therefore a clear contradiction at the core of all these results that needs to be resolved.

Probably the most thoroughly studied transitions are those at 140.3K and 201 K, reported independently by Cazayous et al.[26] and Singh et al.[27] These transitions are manifest in magnon Raman scattering as divergences in cross-section, but they have remained controversial because they do not appear in careful measurements of bulk magnetometry or specific heat, such as those in Figure 1. This has led to speculation that these may be anomalies of extrinsic origin (e.g. second phases, magnetic impurities, or simple artefacts). However, the measurement of Raman magnon linewidth narrowing[26, 27] rules out magnetic impurities, as does the observation of critical exponents for Raman cross-sections[32, 33] and Almedia-Thouless dependence for field-cooled and zero-field-cooled magnetization in thin film samples.[30]

Very recently (2011), two papers have shed additional light on this aspect. Marti et al.[34] have shown, using impedance analysis and grazing incidence x-ray diffraction, that the surface-layer ("skin") of $BiFeO_3$ has a surface-confined phase transition[34], and suggested that some of the cryogenic anomalies of $BiFeO_3$ may also be confined to its skin layer. Meanwhile, Ramazanoglu *et al.*[35] have shown that extremely small uniaxial pressures change the magnetic domain structure strongly, and inferred from that that the low-T transitions (at 140K and 201K) may be linked to such phenomena, which mimic magnetic reorientation transitions like those in orthoferrites. Certainly the fact that the low temperature anomalies tend to be clearer in surface-sensitive probes such as back-scattering Raman experiments would support the idea that these transitions are confined in the surface. At the same time, the strong effect of stress on the magnetic configuration suggests that if the surface is structurally different from the interior, so will its magnetic behaviour. In the present paper, we show that the 140K transition in $BiFeO_3$ is indeed that of a surface phase, and we characterize its structural and electronic properties.

## 2. Raman spectroscopy

As already mentioned, the low-T phase transitions were evidenced using Raman spectroscopy techniques, especially in the low-wavelength region through the analysis of the Magnetic Field Cooling (MFC) versus Zero Field Cooling (MZFC) regime and the study of the electromagnons.[22, 23] The Raman spectrum measured at 80K on a single crystal shows no significant change compared to that at room temperature. However, it is known that any static and/or dynamic changes in the structure should, in principle, lead to a variation in the phonon behaviour, and the analysis of the wave-number, intensity and/or linewidth evolution of the whole spectra as a function of temperature is expected to give insight into those changes. Figure 2 shows typical temperature dependences of the wavelength position for two different Raman bands. Several features are noteworthy. First, all the Raman phonon modes, and not only those related to the electromagnons,[22, 23] show changes in the low temperature regime: whatever the mode, a change of slope occurs at 140K. However, the sign of the slope change (softening vs hardening) is different for different modes. At higher temperature, a new change of slope appears either at 200K or higher or unexpectedly at 180K. As an example, the E-type phonon mode position at around 80cm$^{-1}$ (Figure 2a) is nearly constant at ~83cm$^{-1}$ from 80K to 140K and then continuously increases until 180K, reaching a value of ~97cm$^{-1}$ that remains constant until room temperature. Note that the same behaviour was found in several different samples including single crystals and powders.

## 3. Elasticity

Phonon frequencies are directly linked to inter-atomic forces, so the fact that all the Raman lines shift in the 140-200K range signals that changes in the elastic constants are taking place. In order to test the extent to which these are related to changes in elastic properties, single crystal and ceramic samples have been investigated by Resonant Ultrasound Spectroscopy (RUS).[36] Elastic resonances are dominated by shearing motions and the measured elastic constants scale with $f^2$ (where $f$ are the frequencies of resonance peaks). The inverse mechanical quality factor, usually given as $Q^{-1} = \Delta f/f$, is a measure of anelastic losses associated with the application of a dynamic shear stress. For the present study, RUS spectra in a frequency range 0.1-2MHz were collected in the temperature range ~10-295 K in a helium flow cryostat described by McKnight et al.[37,38] Results for $f$ and $Q^{-1}$ obtained from different resonance peaks are given in Figure 3.

Resonance frequencies, normalized to their value at 300K, all decrease with increasing temperature, consistent with thermal softening of the lattice. A deviation from the linear thermal softening starts to appear around 150 K, with a steep increase (elastic stiffening) between ~175 and ~200 K. An equivalently steep drop back to the baseline occurs between ~225 and ~250 K. The breaks in slope of resonance frequencies of the single crystal sample near 150 and 200 K coincide with breaks in slope of the Raman data (Figure 2).

Frequency data for resonances of the ceramic sample do not show these sharp features, but data for $Q^{-1}$ (elastic losses) from both the ceramic and single crystal samples show similar anomalies in the temperature range of interest: (i) there is a slight break in the slope of the baseline variation in the vicinity of 150 K, from relatively low and fairly constant losses at low temperatures to a trend of increasing loss with increasing temperature; (ii) all the resonances show a peak in $Q^{-1}$ centred on ~180 K, and (iii) there is a further peak or break in slope at ~240 K for the single crystal data, and less well resolved anomalies above ~225 K for the ceramic sample. The break in the slope of $Q^{-1}$ near 150 K is reminiscent of increasing dissipation due to disordering of protons during heating of the mineral lawsonite,[38] though the magnitude of the effect is much smaller. If the analogy is correct, some element of structural or magnetic disordering occurs within the samples above ~150 K. Increasing dissipation implied by the $Q^{-1}$ data could be understood as implying that the structure stable above ~150 K has more disorder (static or dynamic) in comparison with the structure stable at lower temperatures.

All in all, the RUS measurements indicate significant coupling of strain with the changes in structural or magnetic properties identified in other measurements, and suggest the presence of a dissipative –perhaps disordered- state in the temperature range 150K-250K. The measurements, however, do not allow discrimination between phase transitions which occur within the bulk of the sample from one which only occurs within the skin. We nevertheless note that elastic properties probed by sound-propagation measurements (which are only sensitive to bulk as the sound wavelength is of the order of hundreds of microns) show no anomalies at all in this temperature range.[39]

### 4. X-ray diffraction

In order to gain further insight, x-ray diffraction (XRD) was also used. In particular, to discriminate between surface and bulk contributions, we will compare data collected in grazing incidence diffraction (GID) and standard co-planar geometry. The grazing incidence measurements were performed in the ID01 line at the ESRF synchrotron in Grenoble. In contrast to the bulk-sensitive coplanar diffraction, GID allows tuning of the information depth by tuning the incidence angle and/or the photon energy. Following the approach of the preceding high-temperature study,[34] we monitored only the changes in the length of the reciprocal space vectors (modulus of $q$) rather than both their length and direction (vector $q$). This allows for evidence of structural changes confined in skin layers while circumventing the alignment difficulties inherent to single crystals with strong twinning and mosaicity.

Figure 4 (open symbols) shows the relative change of |q| for the (202) reflection measured on a single crystal as a function of temperature in the heating regime from 100K to 300K. The bulk temperature dependence displays no hint of structural change inside the crystal in this temperature range, and only a subtle change in thermal dilatation coefficient from $6.4.10^{-6}$ $K^{-1}$ to $9.4.10^{-6}$ $K^{-1}$ at 180K –which is also the temperature of the first peak in anelastic loss. In contrast, the surface-sensitive data for the (101) peaks reveal an abrupt expansion up to ~1 % between T = 140 K and T ~ 180 K (Figure. 4, solid symbols). This anomaly was not detected in coplanar diffraction where the information depth surpasses few microns. This indicates that the structural change is confined in a surface layer. The surface layer thickness cannot be stated beyond the upper bound placed by the penetration depth of the coplanar geometry (microns), but the sample is nevertheless the same for which a transition at T*=550K was estimated to be within the topmost 10nm. As a matter of fact, it appears that in addition to the phase transition occurring at 550K, the nearby-surface layer overcomes at least another phase transition at ~140K and thus has its own phase diagram.

**5. Impedance analysis and pyroelectric-like currents**

Impedance analysis is an effective tool to probe the electronic properties of surface layers. In particular, it has been noted that Maxwell-Wagner behaviour usually arises whenever there is a substantial difference in conductivity between the bulk and interfacial regions: at low enough frequencies, the contact impedance dominates and the interfacial properties are evidenced[40, 41, 42]. Indeed, this appears to be the case also for our single crystals (Figure 5).

The impedance shows a strong frequency dependence typical of two lossy dielectric components in series.[40,41,42] However, a sudden drop of the impedance (which is equivalent to a sudden increase of the capacitance) is also observed at 140K, which is frequency-independent and thus corresponds to a true phase transition. The fact that the jump in the impedance is bigger for lower frequencies is consistent with the phase transition occurring at the interface of the crystal, in a behaviour analogous to that observed in the interfacial T* transition at 550K.[34] The surface transition appears to be first order, as attested by the sharpness of the jump and by the difference in the critical temperatures on cooling and heating regime (Fig 5 right). This is also consistent with the change in the unit cell volume observed in the grazing incidence diffraction results shown in Figure 4. Recently, Ashok Kumar *et al*. [43] have discovered abrupt onsets of in-plane dielectric loss at 550K and 201K by using interdigital electrodes, which are more sensitive to in-plane surface impedance. This complements our data and supports our interpretation as surface transitions.

In order to test the electronic properties of the surface, we also performed pyroelectric measurements. These are shown in Figure 6. The results show a very sharp and sudden peak in pyroelectric current near 140K. We measured the current discharge both for zero-field-cooling-and-heating and for zero-field-heating-after-field-cooling regimes; in addition to the 140K anomaly, the field-cooled sample shows a further broad anomaly around 200K, plus a sharp jump at 280K. The last anomaly is ascribed to artefact during the measurement. It is worth mentioning that the anomalies at ~140K and ~200K are observed in very many different samples included unprocessed surfaces indicating that those anomalies are intrinsic and not related to chemical etching or mechanical polishing effects.

Unlike in a classical ferroelectric phase transition, where the pyroelectric peak position in temperature does not depend on the poling history of the sample, here the field-cooled sample has the peak at a significant lower temperature (5K less) than the zero-field-cooled one. This shift of peak position toward lower temperatures is a fingerprint of a thermally stimulated current: a current that is generated by emission of trapped charge from a trap level in the forbidden gap of BFO.[44] So, while pyroelectric currents are often related to changes in polarization, here we believe that the current we measured is *not* due to ferroelectricity but to charge injection and thermally stimulated emission from trapping centres. When the skin layer undergoes the phase transition, the Fermi level is likely to experience an abrupt re-arrangement. As a result, interfacial defect states below the Fermi level might cross over above it and release their charge, causing the abrupt thermally-stimulated peak in current. The electronic mechanism for the 140K pyroelectric-like anomaly is also consistent with electron paramagnetic resonance results in large surface-to-volume samples (BiFeO$_3$ nanotubes), discussed in the next section.

### 6. Electron Paraelectric Resonance (EPR) and magnetism in BiFeO$_3$ nanotubes

Because the analyses above emphasize surface phase transitions, it seems useful to prepare samples which maximize the surface to volume ratio. To this aim, we prepared BFO nanotubes and characterized them via EPR, which is sensitive to relatively small volumes. Details of the fabrication process are given in the appendix. The EPR curves are fitted with a Lorentzian line shape when the sample is purely insulator and then the line is perfectly symmetric, or with a Dysonian type function when the line is an asymmetric reflecting conduction component: EPR$_{Dysonian}$ = Absorption×cos($\alpha$) + Dispersion×sin($\alpha$). The asymmetry is described by the parameter $\alpha$ and its value for insulator is zero and 1 for a full conductor.[Erreur ! Source du renvoi introuvable.]

It is clear from the EPR data (Figure 7) that the sample's conductivity is maximum at ~140K and that the conductivity behaviour changes again, less abruptly, at ~200K. Therefore, the EPR data for the nanotubes also indicate an increase in surface charge density at 140K, consistent with detrapping trap levels; the charge released at 140K causes the large pyroelectric-like current observed in Figure 6.

Also relevant to the results here, we note that BiFeO$_3$ samples with large surface to volume ratios (e.g., nanocrystals) have been shown to display spin-glass behaviour.[46] Glassy effects are well known for small particles of antiferromagnets in general, where the lack of spin compensation at the surface is thought to frustrate the long range magnetic order (e.g. NiO[47,48]). Three new EPR observations (Figure 8) are similar to those known in other spin glasses, especially that in $Cd_{1-x}Mn_xTe$:[49] (1) The gyro-magnetic ratio is g > 2 above the apparent spin-glass transition at $T_{sg}$ = 140K and g = approximately 2.0 below; (2) the decrease in g-value at 140 K is rather abrupt with temperature and nearly 1% in magnitude; (3) there is a divergence in EPR linewidth that satisfies a dependence $\Gamma = \Gamma_0 + \Gamma_1 \exp[-T/T_f]$, as shown in Figure 8, with an extrapolated freezing temperature $T_f$ = 33±3 K that is in good agreement with that measured independently as 29.4±0.2 K.[28]

We note also that the EPR susceptibility of the nanotubes is increased between 125 K and 200 K, which is essentially the same temperature range (bearing in mind the sample difference between the nanotubes and the single crystals) where structural disorder has been inferred from elastic spectroscopy. This suggests that the structural disorder has its replica in the magnetic behaviour. We note also that the glassy fitting to the EPR linewidth (Figure 8-middle) departs from the actual data below the skin transition temperature, suggesting a transition from a glassy or magnetically soft state to a more rigidly ordered configuration, the details of which are at this point unknown. We nevertheless emphasize that it is not easy to distinguish magnetoelectric spin-glasses from crystals with domain-wall pinning,[50, 51] both of which would be consistent with the magnetic and elastic results. An extremely fine pattern of domains has in fact been observed in the near-surface region of $BiFeO_3$[52] so this is not out of the question.

### 7. Electronic structure

To gain further insight on the origin of the surface properties, we also performed some first-principle density-functional calculations by introducing some defects which can exist at the surface of BFO samples. Due to bismuth volatility, the most likely defects are Bi vacancies. The calculations have been performed with the SIESTA code.[53,54] Two approximations for the exchange-correlation energy have been tested: the Local Density Approximation (LDA) and the Generalized Gradient Approximation in the form of Perdew, Burke and Ernzerhof (GGA-PBE).[55] Troullier-Martins pseudopotentials have been used. Semicore electrons (3p for Fe, 5d for Bi) are explicitly treated as valence electrons. The equivalent plane wave cutoff for the grid is 200 Ry in the LDA case and 400 Ry in the GGA case. The excitation energy defining the range of the atomic orbitals is 0.01 Ry. The periodic parts of the Kohn-Sham wave functions are expanded on a basis of numerical atomic orbitals of double zeta type (plus polarization orbitals). A single-zeta 7s type orbital is added in the basis set of Bi.

We first optimized bulk BFO in its R3c phase (both in LDA and GGA) and used the lattice constant obtained to build a 2x2x2 supercell (thus containing 80 atoms), whose shape and volume was kept fixed. Then Bi vacancies were introduced in the supercell, in different charge states (0, -1, -2, -3). In the case of charged defects, neutrality is insured by adding a background compensating jellium. Finally the atomic geometries were optimized by using a conjugate-gradient scheme, so as to obtain Cartesian components of atomic forces below 0.04 eV/Angstrom. Figure 9 shows the electronic Density of States (DOS) versus energy for stoichiometric (black curve) and non-stoichiometric BFO supercells containing Bi vacancies.

The calculated LDA energy band-gap for stoichiometric BFO is 0.8 eV, an underestimate compared with the experimental value 2.74 eV;[13] this underestimate is typical of LDA calculations. The main effect of Bi vacancies, whatever the charge of the defect, is to introduce energy levels within the band-gap. These levels can explain the trapping-detrapping process suggested by our electrical measurements.

It is also interesting to remark that Bi vacancies modify the magnetism of the system. Electrons are rearranged giving rise in some cases to a net magnetic moment probably associated with a hole polaron. This rearrangement which can occur within the surface may then explain the concomitant magnetic anomalies observed in the magnon spectra and in the EPR susceptibility. The possibility of the existence of polaron within the close surface is reinforced because this charge-phonon coupling would also explain the anomalies observed in XRD, Raman and RUS data.

## 8. Discussion

The grazing incidence XRD results show unambiguously that the anomaly at 140K corresponds to a surface phase transition. Its key features are an abrupt change in unit cell volume (which expands by 1% on heating, Figure 4), and a concomitant change in electronic structure, with an impurity level crossing the Fermi level and releasing charge, as signalled by the field-dependent pyroelectric-like discharge (Figure 6) and increased conductivity inferred from AC impedance (Figure 5) and EPR analysis (Figure 7). Ab initio calculations show that Bi vacancies might be at the origin of the impurity levels (Figure 9). The Raman spectra also show that the 140K anomaly is strongest in the magnon peaks, and EPR confirms that this transition affects the magnetic structure, as also suggested by the first-principle calculations.

Now the question is what could be the origin of these anomalies. In the perovskite structure, two structural degrees of freedom can be considered; either atomic (polar) displacements or oxygen octahedral tilts. In the case of magnetic materials, as is $BiFeO_3$, a third degree of freedom is the spin. Raman spectroscopy is very sensitive to oxygen octahedral rotations,[56] and yet the number of peaks in the Raman spectra was not observed to change as a function of temperature --though surface-sensitive UV Raman would be desirable to confirm this. The abrupt change in the lattice volume, as indicated by the grazing incidence XRD, points out instead to a change in atomic distances without a change in symmetry. Given that the in-plane lattice parameters of the surface must be coherent with those of the skin, means that a strong uniaxial strain is developed at 140K, which is relevant because uniaxial strain has recently been reported to have a very strong effect on the magnetism of $BiFeO_3$.[35]

The change in unit cell volume is the likely culprit of the crossover of a shallow impurity level across the Fermi line, resulting in charge release.

## 9. Conclusions

The data presented here confirm the interpretation of the 140K anomaly in BiFeO$_3$ as a surface phase transition, with surface effects detected in very many different bulk single crystals and exacerbated in nanotubes owing to their very high surface to volume ratio. The main features of this phase transition are a sharp volume change without actual change of symmetry; sharp emission of charge at 140K (pyroelectric-like current) and maximum in conductivity (peak in the α parameter of the EPR dysonian lineshape), consistent with a crossover between an impurity level and the Fermi level, and structural and magnetic disorder between 140K and c.a. 200K.

As was argued for its high temperature T* counterpart, the surface phase transition at 140K is likely to be aided by the inherent complexity of the phase diagram of BiFeO$_3$, which is very sensitive to even small perturbations[57] such as surface tension or local strain fields around vacancies and defects. A melting of the Bi sublattice has been reported for BiFeO$_3$ powders with a radius smaller than 9nm:[58] if there were Bi vacancies at the surface layer, these would be able to provide both electronic impurity levels and local strain fields capable of explaining the electronic, magnetic and structural changes.

More generally, these results indicate that it is not appropriate in general to treat BiFeO$_3$ as a homogeneous material. Its skin layer is rather different from the bulk, having its own structural, electronic and magnetic properties, and its own phase diagram that already includes at least two confirmed phase transitions at 140K and 550K, as well as a probable glassy state between 140K and 230K. The surface is at least as important as the bulk for functional devices, as it determines key properties such as magnetic exchange bias and conductive barrier height. It is therefore of utmost importance that its nature and properties be fully understood.

## 10. Appendix

*BiFeO$_3$ (BFO) single crystal growth*

BFO single crystals were grown using a method similar to the original method proposed by Kubel and Schmid.[59] Adjusting the cooling rate allows growing of millimeter diameter of rosette-like pyramidal crystals, as described by Burnett et al.[60] All crystals were polished parallel to the surface, which in rosette crystals is the (100) crystallographic plane. Samples typically larger than 1x1 mm$^2$ area and 300 μm thick were obtained. Optical quality crystal surfaces were obtained by polishing using 0.25 μm diamond paste. The remaining damaged surface layer and polishing scratches were removed by chemical mechanical polishing (CMP). CMP was performed usually for 30 min using SiO$_2$ colloidal solution (Syton) diluted with water in a 1:1 ratio.

*BFO nanotube growth*

The nanotubes were prepared via wet-chemistry synthesis: In a typical synthesis, Bi(NO$_3$)$_3$•5H$_2$O was initially added to ethylene glycol to ensure complete dissolution followed by Fe(NO$_3$)$_3$•9H$_2$O to yield a molar ratio in solution of Bi:Fe as 1:1.[39,61] The resulting mixture was stirred at 80 ºC for 1 h, after which a transparent sol was recovered upon evaporation of the excess ethylene glycol. Droplets of the sol were deposited using a syringe onto a porous anodic alumina (AAO) template (Whatman Anodisc®) surface with application of pressure.[40,41] AAO membranes with different pore sizes, such as 200 nm and 100 nm, have been used. The resultant samples of AAO templates containing the BiFeO$_3$ precursors were subsequently oven-dried at 100°C for an hour and then preheated to 400°C for three separate runs at a ramp-rate of 5°C/min in order to get rid of excess hydrocarbons and NO$_x$ impurities. The sample was further annealed at 600°C for 30 min. BiFeO$_3$ nanotubes were isolated after removal of the AAO template, following its immersion in 6M NaOH solution at room temperature for 24 h. Thereafter, to retrieve reasonable quantities of nanotubes, the base solution was diluted in several steps with distilled water and lastly ethanol. Tubes were collected by centrifugation. The tubes were shown to be ferroelectric, with switching hysteresis. The nanotubes were subjected to electrical switching by applying a voltage across a single tube, with the Ir/Pt tip of an atomic force microscope serving as the top electrode. The measured piezoelectric constant hysteresis is quite large (about 2/3 the value of PZT).

*Grazing incident diffraction (GID)*

For the GID experiments (performed at ID01 beamline at ESRF), we chose 7 keV and 0.2 degrees as incidence angles, thereby limiting our information depth to few nanometers. The BiFeO$_3$ crystal was cooled using the Oxford Cryojet blowing cold nitrogen gas on the sample, while temperature was measured in the gas stream and by a thermocouple attached at one side of the crystal (1 mm thick). The crystal employed was the same one as in the GID high temperature study [34].

**Acknowledgements**

Research at Stony Brook and Brookhaven National Laboratory (including support for TJP and SSW as well as for synthesis experiments) was supported by the U.S. Department of Energy, Basic Energy Sciences, Materials Sciences and Engineering Division under contract number DE-AC02-98CH10886. G.C. acknowledges funding from project MAT2010-17771. J.F.S, G. C and M.A thank the Leverhulme trust for supporting their collaboration. The ESRF and the ID01 Beamline staff are acknowledged for their support. X.M. acknowledges "Czech Science Foundation (Project P204/11/P339) ". All the authors want to dedicate this paper in memoriam to their colleague and friend Robert Blinc who unfortunately just passed away.

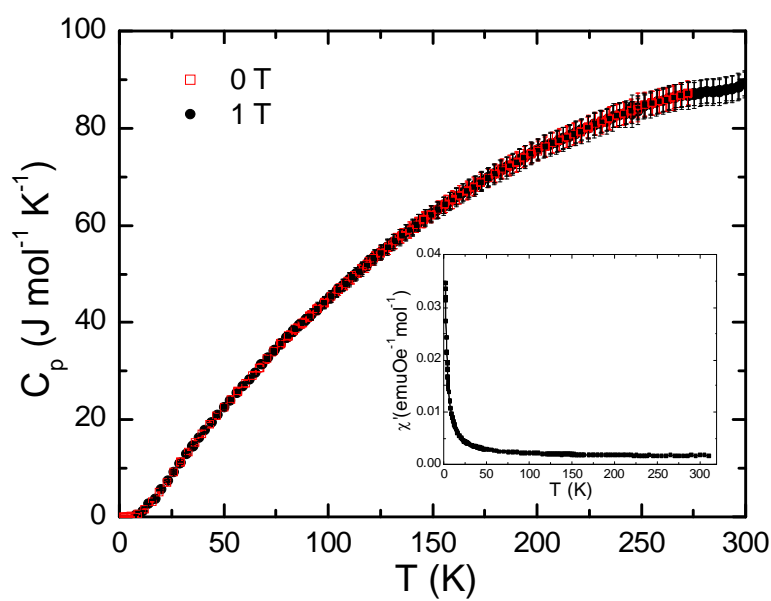

**Figure 1.** Specific heat at 0 and 1 T and ac magnetic susceptibility at 0 T (inset) as a function of temperature for BFO single crystal.

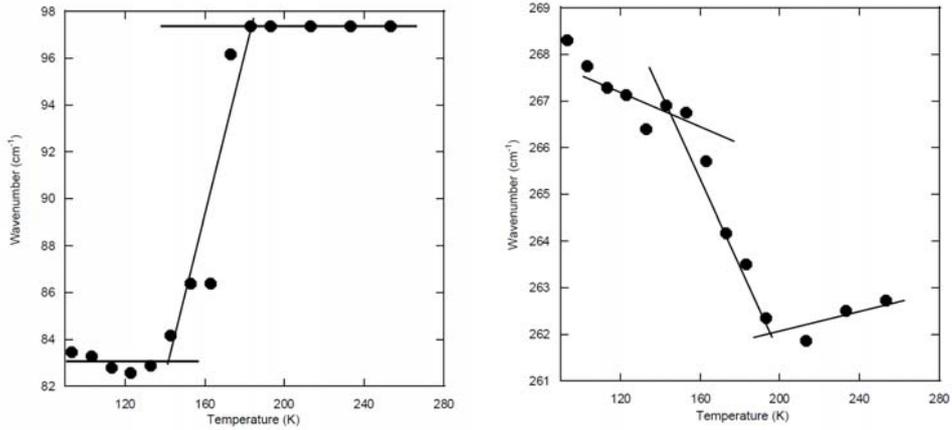

**Figure 2.** : Positions of two Raman peaks as a function of temperature, showing shifts beginning at 140K, with sharp anomalies at 180K, and further changes of slope at 200K.

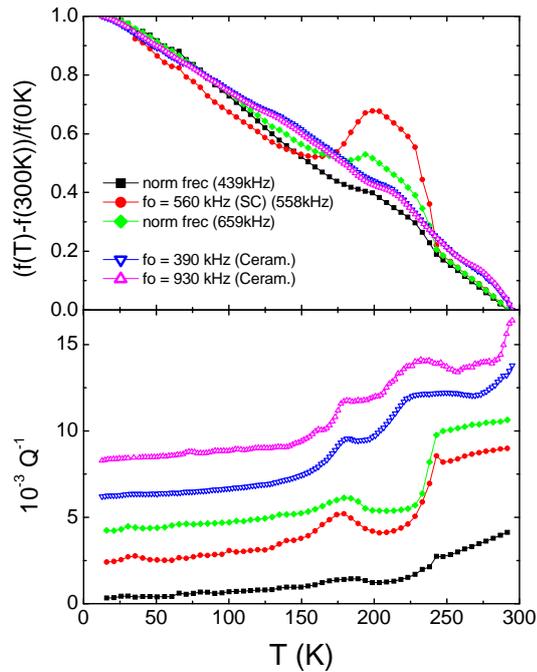

**Figure 3.** (top)RUS results for a BFO single crystal (full symbols) and a ceramic sample (open symbols): the resonant frequencies increase between ~140 and ~240 K, indicating a hardening of the lattice between these two temperatures. (below)The elastic loss (inverse of the quality factor) shows gradual increase above ~150 K, with peaks at ~180 K and at 220-240 K depending on the sample.

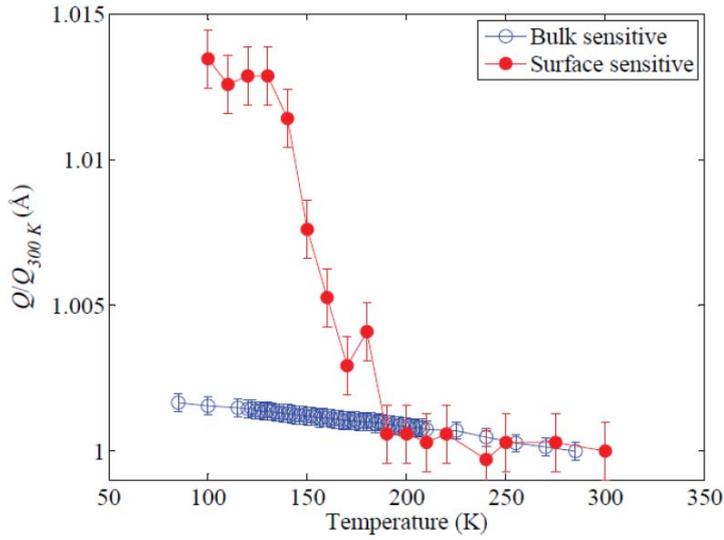

**Figure 4.** Relative change of the reciprocal lattice vector |q| as a function of temperature probing the bulk of the crystal (blue open symbols) and the top-most surface (red solid symbols). The surface data show a rapid expansion of the lattice parameter upon heating above 140K, and this feature is absent from the bulk.

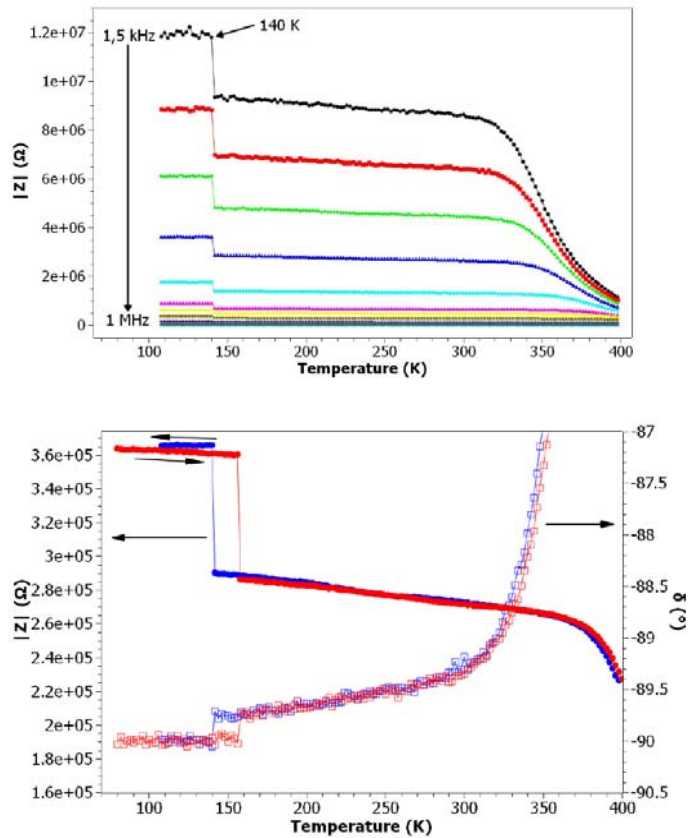

**Figure 5.** (Top) Z modulus vs temperature and frequency; (bottom): Z modulus and impedance phase angle as a function of temperature on heating and cooling showing a first-order phase transition at ~140K.

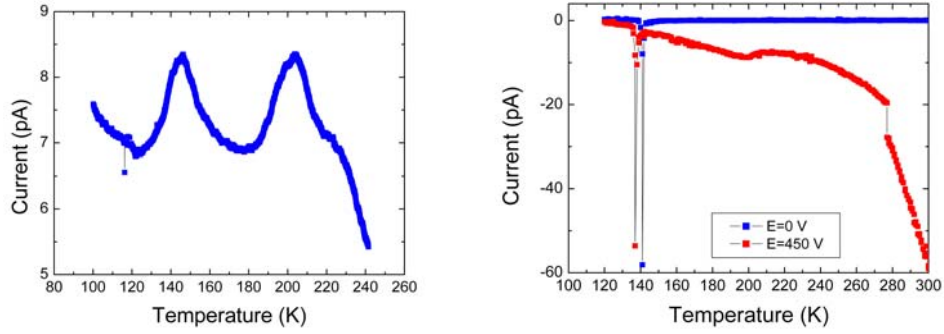

**Figure 6.** Discharge current anomalies in BiFeO$_3$ single crystals. (left) pristine samples show two clear anomalies at ~140K and ~200K, though in subsequent runs (right) only the 140K anomaly is clear, although the 200K anomaly is still visible for field-cooled samples. The field-cooling dependence of the peak temperature for the 140K anomaly indicates that this pyroelectric-like current is due the sudden carrier emission from trap levels triggered by the surface phase transition.

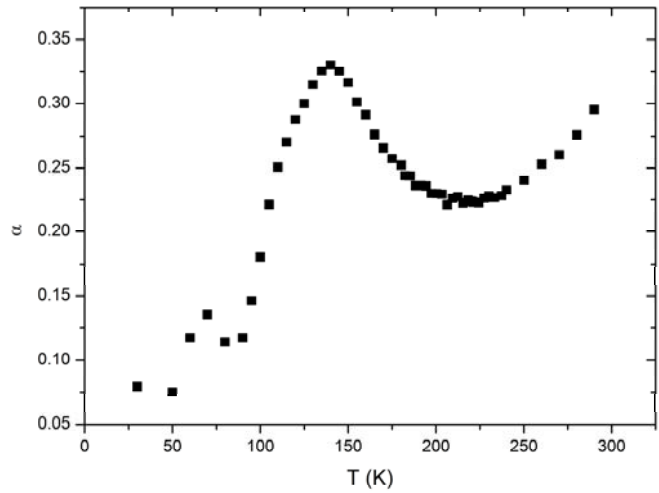

**Figure 7.** Alpha parameter reflecting the asymmetry of the EPR curves.

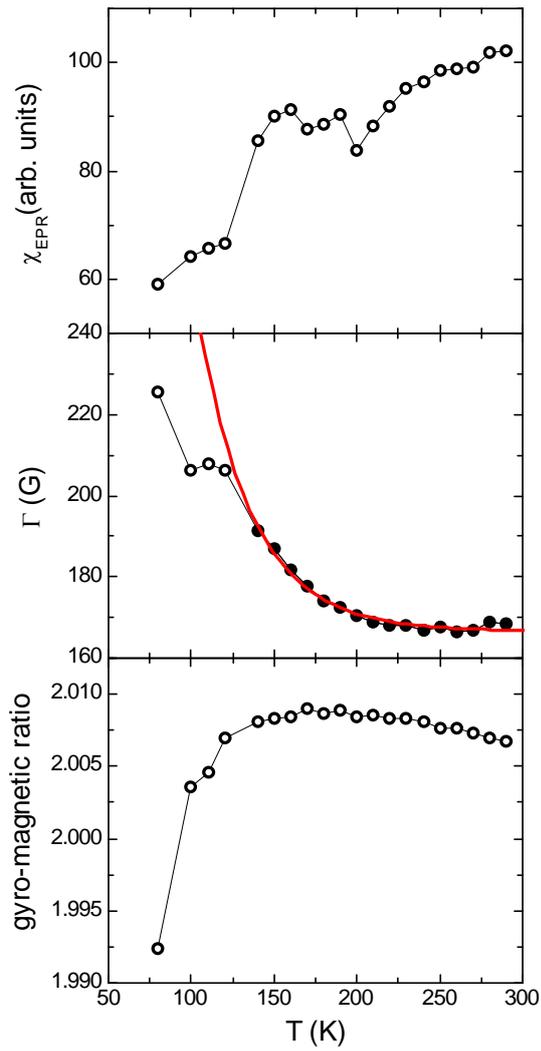

**Figure 8.** (top panel) Electron paramagnetic susceptibility, showing an increase between ∼125K and 200K, which is close to the temperature range where elastic softening and anomalous skin expansion has been detected; (middle panel) an exponential fitting to the EPR linewidth yields an extrapolated freezing temperature of c.a. 33K; the experimental data departs from the glassy fit for temperatures below 140K; (lower panel) the gyromagnetic factor also shows a rather abrupt drop below the skin transition temperature. The EPR susceptibility shows a sharp drop at 201K, whereas the g-value (bottom panel) shows a broad maximum near this temperature.

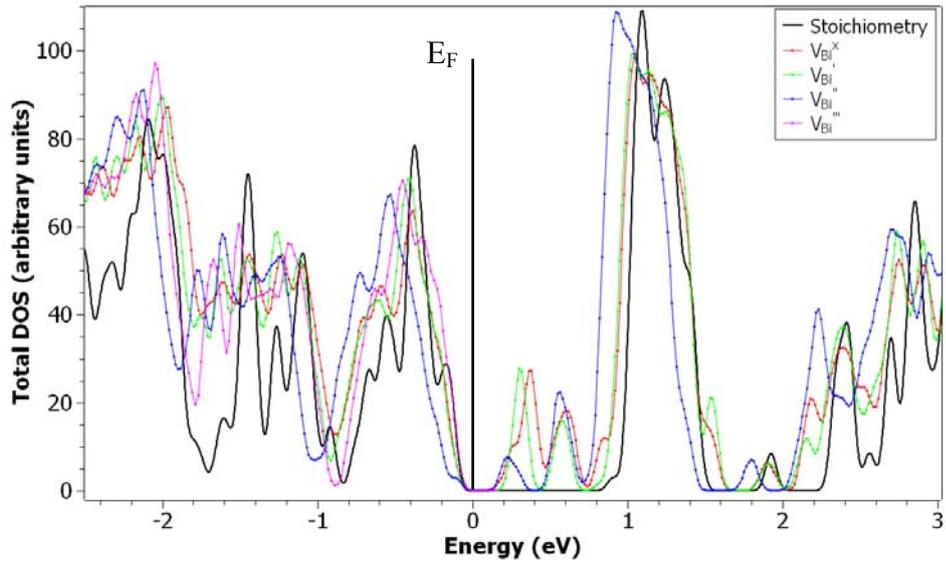

**Figure 9.** Density Of States versus energy from ab initio calculations for pure BFO (black) and BFO with $V_{Bi}$ vacancies in different charge states ($V_{Bi}^{x}$, $V_{Bi}'$, $V_{Bi}''$, $V_{Bi}'''$).